\documentstyle[preprint,aps,epsfig,eqsecnum]{revtex}

\begin{document}  

\tightenlines


\draft


\title{Limits on Majorana Neutrinos from Recent Experimental Data  }

\author{ O.\ L.\ G.\ Peres$^1$\thanks{Email address: operes@flamenco.ific.uv.es},
L.\ P.\ Freitas$^2$\thanks{Email address: lfreitas@charme.if.usp.br},
and R.\ Zukanovich Funchal$^2$\thanks{Email address: zukanov@charme.if.usp.br}}

\address{\em $^1$ Instituto de F\'{\i}sica Corpuscular -- C.S.I.C. \\
    Departamento de F\'{\i}sica Te\`orica, Universitat of Val\`encia \\
    46100 Burjassot, Val\`encia, Spain\\[8pt]
    $^2$  Instituto de F\'\i sica, Universidade de S\~ao Paulo\\
    C.\ P.\ 66.318, 05315-970 S\~ao Paulo, Brazil}

\maketitle
\vspace{.5cm}

\hfuzz=25pt
 
\begin{abstract}

We investigate the sensitivity of some weak processes to the simplest
extension of the Standard Model with Majorana neutrinos mixing in the 
leptonic sector. Values for mixing angles and masses compatible with 
several experimental accelerator data and the most recent neutrinoless 
double-$\beta$ decay  limit were found.

\end{abstract}

\newpage

\section{Introduction} 
\label{sec:int}

Experimental neutrino physics has regained great interest in the latest 
years, with many new experiments presently taking data or in
preparation for the near future. This is justified because although
the Standard Model has been  vigorously tested experimentally and  
seems to be a remarkably successful description of nature,  
its neutrino sector has yet been poorly scrutinized.  
We believe that this still mysterious
area of particle physics may give us some hint on the physics beyond
the Standard Model.

It is a common prejudice in the literature to assume 
the conservation of the leptonic number and  to think about 
neutrinos as Dirac particles much lighter than any of the charged 
leptons we know. Nevertheless there are no theoretically compelling 
reasons why the leptonic number should be a conserved quantity or 
why neutrinos should not have a mass comparable to the charged fermions.
It is clear that only the confrontation of theory 
with experimental data will eventually clarify the problem of 
neutrino mass and nature.  

Many direct limits on neutrino mass have been obtained by different
experimental groups~\cite{pdg98} but are not all accepted without
controversy~\cite{tau4,pdg}.  Experiments also have been carried out
to try to measure neutrinoless  double-$\beta$ decay which, in general, 
is a process that will not occur unless one has  a Majorana neutrino
involved as an intermediate particle. Here also experiments have
obtained only limits on the so called effective neutrino mass~\cite{hmbb}. 
As a rule experimental analysis are model dependent and cannot be  
quoted as a general result.

In the hope of contributing to the understanding of neutrinos physics
we have accomplished a comprehensive study of the constraints imposed
by recent experimental data on lepton decays, pion and kaon leptonic
decays as well as by the $Z^0$ invisible width 
measurement performed by the LEP experiments to the simplest model
containing Majorana neutrinos.  

We will consider a very simple extension of the standard electroweak
model which consists in adding to its particle content a right-handed 
neutrino transforming as a singlet under $\text{SU(2)}_L\otimes \text{U(1)}_Y $. 
This will be referred as the Minimal Model with Right-handed Neutrino (MMRN). Next,
by allowing it to mix with all the left-handed neutrinos we obtain
that there are, at the tree level, two massless neutrinos ($m_1$, $m_2$)
and two massive ones ($m_P$, $m_F$)~\cite{cj}.

It is interesting to note that this simple extension of the Standard
Model imposes a mass hierarchy for neutrinos.  
The massless neutrinos ($m_1$, $m_2$) can acquire
very small mass by radiative corrections~\cite{babu,gandhi}. This seems to 
be consistent with the recent evaluation of the the number of light 
neutrino species from big bang nucleosynthesis~\cite{nucs}.

The outline of this work is as follows. In Sec.~\ref{sec:mod} the
model consider is briefly reviewed. In Sec.~\ref{sec:lepmix}
we consider the effects of mixing for the decay width of the muon, 
for the partial leptonic decay widths of the tau, pion and kaon 
and for the $Z^0$ invisible width. These are the quantities that are calculated
theoretically.  In Sec.~\ref{sec:ana} we compare our theoretical
results with recent experimental data and obtain from this comparison
allowed regions for mixing angles and masses.  In Sec.~\ref{sec:dbeta}
we investigate the possibility of further constraining our results with 
the present best limit from neutrinoless double-$\beta$ decay experiments. 
Finally, in the last section we establish our conclusions.

\section{A brief description of the model}
\label{sec:mod}

In the MMRN the most general form of the neutrino mass term is

\begin{equation}
{\cal L}_{\nu}^{M}
=-\sum_{\alpha=e,\mu,\tau} a_\alpha \overline{{\nu_{\alpha}}}_L
N_{R}-\frac{1}{2}M \overline{N^{c}_{R}}   N_{R}+H.c.
\label{mass}
\end{equation}
where the left-handed neutrino fields are the usual flavor eigenstates and we 
have assumed that the charged leptons have already been diagonalized.  In this
model, there are four physical neutrinos   $\nu_1,\nu_2,\nu_P$ and
$\nu_F$, the first two are massless ($m_1=m_2=0$) and the last 
two  are massive Majorana neutrinos with masses  

\begin{equation}  
m_P=\frac{1}{2}(\sqrt{M^2+4a^2}-M)~ \text{ and } ~ m_F=\frac{1}{2}(\sqrt{M^2+4a^2}+M),  
\label{mass2}  
\end{equation}  
where $a^2=a_e^2+a_\mu^2+a_\tau^2$.

In terms of the physical fields the charged current interactions are
  
\begin{equation}  
{\cal L}^{CC}= \frac{g}{\sqrt{2}}
\begin{array}{cccc}(\overline{\nu_{1}}& \overline{\nu_{2}}& \overline{\nu_{P}}&
\overline{\nu_{F}}\end{array})_L\gamma^\mu \Phi R  
\left(\begin{array}{c}  
e\\ \mu\\ \tau \\ 0 \end{array}\right)_LW_\mu^++H.c.,  
\label{cc}  
\end{equation}  
where $\Phi=diag(1,1,i,1)$ and $R$ is the matrix  
\begin{equation}  
\left(  
\begin{array}{cccc}  
R_{e1} & R_{\mu 1}& R_{\tau 1} & R_{0 1} \\   R_{e2} & R_{\mu 2} &
R_{\tau 2} & R_{0 2} \\   R_{eP} & R_{\mu P} & R_{\tau P} & R_{0 P}
\\   R_{eF} & R_{\mu F} & R_{\tau F} & R_{0 F} \\  
\end{array}\right)  
=\left(  
\begin{array}{cccc}  
c_\beta &-s_\beta s_\gamma & -s_\beta c_\gamma & 0 \\   0 & c_\gamma &
-s_\gamma & 0 \\   c_\alpha s_\beta\; & c_\alpha c_\beta s_\gamma\;
&c_\alpha c_\beta   c_\gamma\; &-s_\alpha \\   s_\alpha s_\beta \;&
s_\alpha c_\beta s_\gamma\; & s_\alpha c_\beta   c_\gamma\; &
c_\alpha  
\end{array}\right).  
\label{mm}  
\end{equation}  
In Eq.~(\ref{mm}) $c$ and $s$ denote the cosine and the sine of the  
respective arguments. The angles $\alpha,\beta$ and $\gamma$ lie in  
the first quadrant and are related to the mass parameter as follows  
\begin{equation}  
s_\alpha=\sqrt{m_P/(m_P+m_F)},  
\label{angles}  
\end{equation}  
\begin{equation}  
s_\beta=a_e/a,\;c_\beta s_\gamma=a_\mu/a,\;c_\beta
c_\gamma=a_\tau/a.  
\label{angles2}  
\end{equation}  
The choice of parameterization is such that for
$\alpha=\beta=\gamma=0$ , $\nu_1\rightarrow\nu'_e$,
$\nu_2\rightarrow\nu'_\mu$ and $\nu_P\rightarrow\nu'_\tau$.

The neutral current interactions for neutrinos written in the
physical basis of MMRN read
\begin{equation}
{\cal L}^{NC}= \frac{g}{4 \cos\theta_W} \left(\begin{array}{cccc}  
\overline{\nu_{1}}&\overline{\nu_{2}}&\overline{\nu_{P}}&\overline{\nu_{F}}\end{array}\right)_L  
\gamma^\mu\left(\begin{array}{cccc}  
1 \;& 0\; & 0 & 0 \\   0 \;& 1\; & 0 & 0 \\   0 \;& 0 \;& c^2_\alpha &
ic_\alpha s_\alpha \\   0 \;& 0 \;& -ic_\alpha s_\alpha &
s^2_\alpha\end{array}\right)  
\left(\begin{array}{c}  
\nu_{1}\\ \nu_{2} \\ \nu_{P} \\ \nu_{F}\end{array}\right)_L Z_\mu+H.c.  
\label{nc}  
\end{equation}  

Notice that there are four independent parameters in MMRN. 
We will choose them to be the angles $\beta$ and $\gamma$
and the two Majorana masses $m_P$ and $m_F$. 
These are the parameters that we will constrain with experimental data.

\section{Four generation mixing in the leptonic sector}  
\label{sec:lepmix}

In this section we will present the expressions that will be used in
our analysis for muon and tau leptonic decays, pion and kaon leptonic decays 
and the $Z^0$ invisible width. The coupling constant $G$ and the decay
constants $F_\pi$ and $F_K$ used in our theoretical expressions have
not the same values of the standard $G_\mu$, $f_\pi$ and $f_K$ given
in Ref.~\cite{pdg98}, this important point will be discussed at the
end of this section.
 
\subsection{Lepton decays}
\label{sec:ldec}

We can now write the most general expression for the partial decay
width of a lepton $l'$ into a lepton $l$ and two neutrinos $\bar \nu_{l}
\nu_{l'}$ in the context of MMRN as

\begin{eqnarray}
\Gamma(l'\to l\bar\nu_l\nu_{l'})&=&\frac{G^2m^5_{l'}}{192\pi^3} 
{\cal R}^{l'} \left\{
\left(\vert R_{l'1}\vert^2+\vert R_{l'2}\vert^2\right)
\left(\vert R_{l1}\vert^2+\vert
R_{l2}\vert^2\right)\,\Gamma^{l'l}_{11}\right.\nonumber \\ &+&   
\left.\left(\vert R_{lP}\vert^2[\vert R_{l'1}\vert^2+\vert R_{l'2}\vert^2]  
+\vert R_{l'P}\vert^2[\vert R_{l1}\vert^2+\vert R_{l2}\vert^2]  
\right)\,\Gamma^{l'l}_{1P}\right.\nonumber \\   
&+&   
\left.\left(\vert R_{lF}\vert^2[\vert R_{l'1}\vert^2+\vert R_{l'2}\vert^2]  
+\vert R_{l'F}\vert^2[\vert R_{l1}\vert^2+\vert R_{l2}\vert^2]  
\right)\,\Gamma^{l'l}_{1F}\right.\nonumber \\   
&+&
\left.\left(\vert R_{lF}\vert^2 \vert R_{l'P}\vert^2+
\vert R_{l'F}\vert^2 \vert R_{lP}\vert^2
\right)\,\overline{\Gamma}^{l'l}_{PF}\right.\nonumber \\ 
&+&\left.\vert R_{l'P}\vert^2 \vert
R_{lP}\vert^2\,\overline{\Gamma}^{l'l}_{PP} + \vert R_{l'F}\vert^2
\vert R_{lF}\vert^2\,\overline{\Gamma}^{l'l}_{FF} \right\},   
\label{decay}  
\end{eqnarray}  
with $l'=\mu,\tau$ and $l=e,\mu$ for the tau decays and $l=e$ for
the muon decay. Notice that $G^2$ in Eq.\ (\ref{decay}) is the 
universal constant defined as $G^2/\sqrt2=g^2/8m^2_W$.

In Eq.\ (\ref{decay}) we have used the integrals

\begin{equation}  
\Gamma^{l'l}_{11}=2\int^{t_M}_{t_m}(t^2-B)^{\frac{1}{2}}[t(3k-2t)-B]dt,  
\label{r0}  
\end{equation}  
\begin{eqnarray}  
\Gamma^{l'l}_{1J} & = &  
2\int^{t_M}_{t_m}(t^2-B)^{\frac{1}{2}}\frac{(k-\delta_{J
l'}^2-t)}{(k-t)^3}   
\left[(k-\delta_{J l'}^2-t)^2t(k-t) \right. \nonumber \\   
 & + &\left. [(k-t)^2+\delta_{J l'}^2(k-t)-2\delta_{J l'}^4]
(2kt-t^2-B)\right] \theta(m_{l'}-m_l-m_J) dt,
\label{r3}  
\end{eqnarray}

\begin{equation}
\overline{\Gamma}^{l'l}_{JJ'} = \Gamma^{l'l}_{JJ'} + \epsilon_{JJ'} {\Gamma}'^{l'l}_{JJ'},\quad \epsilon_{JJ'}= \left\{ \begin{array}{rl}  1 & (J=J') \\
-1 & (J\neq J') \end{array} \right.
\label{baradd}
\end{equation}

\begin{eqnarray}
\Gamma^{l'l}_{JJ'} & = &  
2\int^{t_M}_{t_m}(t^2-B)^{\frac{1}{2}}C_{JJ'}
\left[2(k-t)tC^2_{JJ'}\right. \nonumber \\ &+&\left.(2kt-t^2-B)B_{JJ'}
\right]
\theta((m_{l'}-m_l)-m_J-m_{J'}) dt, \label{rFP}
\end{eqnarray}  

\begin{eqnarray}
{\Gamma'}^{l'l}_{JJ'} & = &
-12\int^{t_M}_{t_m}(t^2-B)^{\frac{1}{2}}C_{JJ'}\delta_{J l'}\delta_{J'
l'}
\theta((m_{l'}-m_l)-m_J-m_{J'}) dt, \label{rFPL}
\end{eqnarray}

with     
\begin{equation}  
k=1+\delta^2_{ll'},\quad B=4(k-1),\quad  
\delta_{Jl'}=\frac{m_{J}}{m_{l'}},\quad  
\delta_{ll'}=\frac{m_l}{m_{l'}},    
\label{def1}  
\end{equation}  
  
\begin{equation}  
t_m=2\delta_{ll'},\quad   t_M=k-\frac{(m_{i}+m_{j})^2}{m^2_{l'}},   
\label{def2}  
\end{equation}  

\begin{equation}
C_{JJ'}=\frac{\left[
(k-t)^2+(\delta_{Jl'}^2-\delta_{J'l'}^2)^2-2(\delta_{Jl'}^2+\delta_{J'l'}^2)(k-t)\right]^\frac{1}{2}}{k-t},
\label{cjj}
\end{equation}

\begin{equation}
B_{JJ'}=\frac{2}{(k-t)^2}\left[
(k-t)^2-2(\delta_{Jl'}^2-\delta_{J'l'}^2)^2+(\delta_{Jl'}^2
+\delta_{J'l'}^2)(k-t)\right],
\label{bjj}
\end{equation}
where $i,j=1,2,P,F$; $J,J'=P,F$; $m_{l(l')}$ are the corresponding lepton 
masses; $\Gamma^{l'l}_{11}$ and
$\Gamma^{l'l}_{1J}$ are respectively the phase space 
contributions to the $l'\to l\bar\nu_{l}\nu_{l'}$ 
decays for two massless and one massive
neutrino (for either Dirac or Majorana type neutrinos)~\cite{sharma}.  
If the final state neutrinos were two
massive Dirac neutrinos the contribution would be simply
$\Gamma^{l'l}_{JJ'}$, but since here they are Majorana neutrinos 
there is an additional contribution ${\Gamma'} ^{l'l}_{JJ'}$.
The quantity ${\cal R}^{l'}$ describes the leading radiative
corrections to the lepton decay process that can be found in the
Appendix.

Explicitly using the parameterization given in Eq.\ (\ref{mm}) and defining
$x=s_\beta^2$, $y=s_\gamma^2$ and $z=s_\alpha^2$ we obtain
\begin{equation}  
\Gamma(\mu\to e\nu_\mu\bar\nu_e)=\Gamma^{\mu e}=\frac{G^2m^5_\mu}{192\pi^3}{\cal R}^{\mu}
f^{\mu e}(x,y,\delta_{e\mu},\delta_{P \mu},\delta_{F \mu}),  
\label{def3}  
\end{equation}  
for the partial rate of the muon decay into electron, and 

\begin{equation}  
\Gamma(\tau\to e\nu_\tau\bar\nu_e)=\Gamma^{\tau e}=\frac{G^2m^5_\tau}{192\pi^3}
{\cal R}^{\tau}f^{\tau   e}(x,y,\delta_{e\tau},\delta_{P
\tau},\delta_{F \tau}),  
\label{def4}  
\end{equation}  

\begin{equation}  
\Gamma(\tau\to \mu\nu_\tau\bar\nu_\mu)=\Gamma^{\tau \mu}=\frac{G^2m^5_\tau}{192\pi^3}
{\cal R}^{\tau}f^{\tau  
\mu}(x,y,\delta_{\mu\tau},\delta_{P\tau},\delta_{F\tau}),  
\label{def41}  
\end{equation}  
for the partial widths of the tau decay into electron and muon, respectively. 

The following definitions were used 

\begin{eqnarray}
f^{\mu e}(x,y,\delta_{e\mu},\delta_{P \mu},\delta_{F \mu})
&=&\left[ (xy+(1-y))(1-x)\,\Gamma^{\mu
e}_{11} \right . \nonumber \\ &+& \left.
(1-z)(x^2y+x(1-y)+(1-x)^2y)\Gamma^{\mu e}_{1P}\right.\nonumber \\   
&+&
\left. z(x^2y+
x(1-y)+(1-x)^2y)\Gamma^{\mu e}_{1F} \right.\nonumber \\  &+& \left.
2((1-z)xz(1-x)y)\overline{\Gamma}^{\mu e}_{PF} \right.\nonumber \\ 
&+& \left.  (1-z)^2y(1-x)x\,\overline{\Gamma}^{\mu e}_{PP}
+z^2y(1-x)x\,\overline{\Gamma}^{\mu e}_{FF}
\right],
\label{mue}  
\end{eqnarray}  

\begin{eqnarray}  
f^{\tau   e}(x,y,\delta_{e\tau},\delta_{P
\tau},\delta_{F \tau})&=&\left[(  
x(1-y)+y)(1-x)\,\Gamma^{\tau e}_{11}\right.\nonumber \\ &+& \left.
(1-z)(x^2(1-y)+xy+(1-x)^2(1-y))\,\Gamma^{\tau e}_{1P}\right.\nonumber
\\ &+&\left. z(x^2(1-y)+xy+(1-x)^2(1-y))\,\Gamma^{\tau
e}_{1F}\right.\nonumber \\
&+&\left.2((1-z)(1-x)(1-y)zx)\,\overline{\Gamma}^{\tau
e}_{PF}\right.\nonumber \\
&+&\left.(1-y)(1-x)x(1-z)^2\,\overline{\Gamma}^{\tau e}_{PP}
+(1-y)(1-x)xz^2\,\overline{\Gamma}^{\tau e}_{FF}  
\right],  
\label{taue}  
\end{eqnarray}  

\begin{eqnarray}  
f^{\tau  
\mu}(x,y,\delta_{\mu\tau},\delta_{P\tau},\delta_{F\tau}) 
&=&\left[(x(1-y)+y)(xy+(1-y))\,  
\Gamma^{\tau\mu}_{11}  
\right.\nonumber  
\\ &+&\left. \left(y(x(1-y)+  
y)+(1-y)(xy+(1-y)) \right)(1-z)(1-x)\,  
\Gamma^{\tau\mu}_{1P}  
\right.\nonumber  
\\ &+&\left. \left(y(x(1-y)+  
y)+(1-y)(xy+(1-y)) \right)z(1-x)\,  
\Gamma^{\tau\mu}_{1F}  
\right.\nonumber  
\\ &+&\left. 2 \left( z(1-x)^2y(1-z)(1-y) \right)\,
\overline{\Gamma}^{\tau\mu}_{PF} 
\right.\nonumber  
\\ &+&\left. 
(1-y) y (1-x)^2 (1-z)^2\,\overline{\Gamma}^{\tau\mu}_{PP} +(1-y)y
(1-x)^2z^2\,\overline{\Gamma}^{\tau\mu}_{FF}\right].
\label{taumu}  
\end{eqnarray}  
  
\subsection{Pion and Kaon leptonic decays}
\label{sec:hdec}

We will also consider decays such as $h\to l+\nu_l$; where 
$h=\pi , K ~\mbox{ and } ~l=e,\mu$.

The partial width for the leptonic decay of hadrons in MMRN is

\begin{eqnarray}
\Gamma(h \rightarrow l \nu_l) &=& \Gamma^{h l}\nonumber\\
&=& \frac{G^2F^2_h V^2_{KM} m^3_h}{8\pi} {\cal R}_{hl} f^{hl}(x,y,\delta_{h
l},\delta_{P l},\delta_{F l}),
\label{hadw}
\end{eqnarray}
with  $m_h$ being the mass of the hadron $h$ and 
 
\begin{eqnarray}
f^{hl}(x,y,\delta_{h l},\delta_{P l},\delta_{F l})
&=&\left[ \left(\vert R_{l1}\vert^2 +\vert R_{l2}\vert^2\right)\Gamma^{h l}_1+
\vert R_{lP}\vert^2 \Gamma^{h l}_P+ \vert R_{lF}\vert^2 \Gamma^{h l}_F 
\right], 
\label{hadwdef}
\end{eqnarray}
where $\Gamma^{h l}_1$ is the massless neutrino contribution given by

\begin{equation}
\Gamma^{h l}_1=(\delta_{hl}^2-\delta_{hl}^4) \lambda^\frac{1}{2}(1,\delta_{hl}^2,0),
\label{hmassl}
\end{equation}
and $\Gamma^{h l}_J$ are the massive neutrino contributions~\cite{ro}
\begin{equation}  
\Gamma^{h l}_J=
\left[\delta^2_{hl}+\delta^2_{Jl}-(\delta^2_{hl}-\delta^2_{Jl})^2\right]  
\lambda^{\frac{1}{2}}(1,\delta^2_{lh},\delta^2_{Jl})\theta(m_h-m_l-m_J),
\label{piw}
\end{equation}
$J=P,F$, $\delta_{hl}=m_l/m_h$, $V^2_{KM}$ is the appropriate
Cabibbo-Kobayashi-Maskawa matrix element of the quark sector and
$\lambda$ is the triangular function defined by
\[\lambda(a,b,c)=a^2+b^2+c^2-2(ab+ac+bc).\]

The quantity ${\cal R}_{hl}$ in Eq.(\ref{hadw}) represents the leading
radiative corrections to the hadron $h$ decay given in the Appendix. 

In particular when the final state is a muon we have
\begin{eqnarray}  
f^{h\mu}(x,y,\delta_{h \mu},\delta_{P \mu},\delta_{F \mu}) &=&\left(\vert  
R_{\mu1}\vert^2+ \vert R_{\mu2}\vert^2 \right)\Gamma^{h\mu}_1+  
\vert R_{\mu P}\vert^2\Gamma^{h\mu}_P +\vert R_{\mu F}\vert^2\Gamma^{h\mu}_F  \nonumber \\&=&  
\left(yx+1-y \right)\Gamma^{h\mu}_1+  
y(1-x)(1-z)
\Gamma^{h\mu}_P+ y(1-x)z
\Gamma^{h\mu}_F,
\label{pimu}  
\end{eqnarray}  
and when the final state is an electron
\begin{eqnarray}  
f^{h e}(x,y,\delta_{h e},\delta_{P e},\delta_{F e}) &=&\left( \vert  
R_{e1}\vert^2 +\vert R_{e2}\vert^2\right)\Gamma^{h e}_1+  
\vert R_{eP}\vert^2 \Gamma^{h e}_P+ \vert R_{eF}\vert^2 \Gamma^{h e}_F
\nonumber \\ &=&   
(1-x)\Gamma^{h e}_1+ (1-z)x \Gamma^{h e}_P +zx \Gamma^{h e}_F.
\label{pie}  
\end{eqnarray}  

\subsection{$Z^0$ invisible width}
\label{sec:Zinv}

In this section we will extend and update our previous analysis in
Ref.~\cite{cec}.  In the MMRN scheme the $Z^0$ partial invisible width can
be written as~\cite{cj}

\begin{equation}
\Gamma^{\mbox{\scriptsize inv}}(Z\rightarrow  
\nu's)=\Gamma_0 (2+( 1-z^2) \chi_{PP}+2 ( 1-z ) z \chi_{PF}
+ z^2 \chi_{FF}),  
\label{width}  
\end{equation}  
where $\Gamma_0$ is given by
\begin{equation}
\Gamma_0=\frac{G M^3_{Z}}{6\sqrt{2}\pi}( \bar g^2_V + \bar g^2_A),
\end{equation}
and the electroweak corrections to the width are incorporated in the 
couplings $\bar g_V$ and $\bar g_A$,

\begin{equation}
\chi_{ij}=\frac{\sqrt{\lambda (M^2_Z,m^2_i,m^2_j)}}{M^2_Z} X_{ij}\theta(M_Z-m_i-m_j),
\label{chi}
\end{equation}
here $i,j=P,F$; $\lambda$ is the usual triangular function already
defined and $X_{ij}$ include the mass dependence of the matrix
elements.   Explicitly,  
\[X_{PP}=1-4\frac{m^2_P}{M^2_Z}, \]
\[X_{FF}=1-4\frac{m^2_F}{M^2_Z}, \]
\begin{equation}
X_{FP}=1-\frac{\Delta m^2_{FP}}{2M^2_Z}-\frac{m^2_P+3m_Fm_P}{M^2_Z}  
-\frac{(\Delta m^2_{FP})^2}{4M^4_Z},
\label{x}  
\end{equation}
where we have defined $\Delta m^2_{FP}=m^2_F-m^2_P$.  Thus,
$\chi_{ij}$ are bounded by unity whereby
\begin{equation}
\Gamma^{\mbox{\scriptsize inv}}(Z\rightarrow \nu's)\leq 3\Gamma_0.
\label{width2}
\end{equation}

\subsection{Comment on $G$ and $F_h$}
\label{sec:const}

It is common to assume that
standard processes will practically not be affected, at tree level, 
by the introduction of new physics, and that the most effective way 
of constraining new physics is by looking at exotic processes. 
This is correct in most situations envisaged in the literature.
For instance in Ref.~\cite{ng} the
emphasis is given to lepton flavor violation processes like
$\mu\rightarrow e\gamma$. 
Nevertheless we would like to point out that constants 
used in the standard weak decays may take different values as a consequence 
of mixing. 

The experimental value for the muon decay constant, $G_\mu$, is
obtained by comparing the Standard Model formula for the muon decay width

\begin{equation}
\Gamma^{\text{SM}}(\mu\to e\bar\nu_e\nu_{\mu})=\frac{G^2_{\mu}m^5_{\mu}}{192\pi^3} 
{\cal R}^{\mu} \Gamma_{11}^{\mu e},
\label{gu}
\end{equation}
with the measured muon lifetime. As the error obtained in this way is 
very small, $G_\mu$ is often used as an input in the calculations of radiative 
corrections~\cite{franchioti}. 

Now if we have mixing the expression for the 
muon decay width is modified as in Eq.~(\ref{def3}). So that  
comparing this equation with Eq.~(\ref{gu}), it is clear that
the numerical value of $G_\mu$ is not equal to the numerical value of $G$, 
as a general rule, independently of the accuracy of $G_\mu$ determination. 
They are related by:

\begin{equation}
G^2 = \frac{\Gamma_{11}^{\mu e} G_\mu^2}{ f^{\mu
e}(x,y,\delta_{e\mu},\delta_{P \mu},\delta_{F \mu})}.
\label{gmug}
\end{equation}

From Eqs.~(\ref{mue}) and (\ref{gmug}) we see that $G \geq G_\mu$.
A consequence of this is that the $Z^0$ invisible decay width 
\begin{equation}
\Gamma^{\mbox{\scriptsize inv}}(Z\rightarrow \nu's) \leq  3\Gamma_0 
= 3 \frac{G}{G_\mu} \Gamma_0^{\mbox{\scriptsize SM}},
\label{Zlim}
\end{equation}
could, in principle, even exceed $3~ \Gamma_0^{\mbox{\scriptsize SM}}$,
where $\Gamma_0^{\mbox{\scriptsize SM}}$ is the Standard Model width. 

In a similar way the experimental value of the pseudoscalar meson decay 
constant $f_h$ is obtained by 
comparing  the Standard Model prediction for the hadron 
leptonic decay width

\begin{equation}
\Gamma^{\text{SM}}(h \rightarrow l \nu_l) = \frac{G^2_{\mu}f^2_h V^2_{KM}m^3_h}{8\pi} 
{\cal R}_{hl} \Gamma^{h l}_1,
\label{hadw_sm}
\end{equation} 
with experimental data.
The values of $f_h$ quoted in PDG depend on the type of 
radiative corrections used~\cite{finke1,sirlin}. The extracted values  
$f_\pi=130.7\pm0.4$ MeV and  
$f_K=159.8\pm1.5$ MeV~\cite{pdg98},  
were obtained using the expression of 
${\cal R}_{hl}$ as in our  Appendix.

Here also the numerical values of $F_\pi$ and $F_K$ are not equal to  
the numerical values of $f_\pi$ and $f_K$ given above, since 
the constant $F_h$ that appears in Eq.~(\ref{hadw}) is 
related to $f_h$ in Eq.~(\ref{hadw_sm}) by

\begin{equation}
\Gamma_1^{h \mu} G^2_\mu f^2_h = G^2 F^2_h f^{h \mu}(x,y,\delta_{h \mu},
\delta_{P l},\delta_{F l}).
\label{Ffhad}
\end{equation}

\section{Experimental constraints on mixing angles and neutrino masses}  
\label{sec:ana}

As we explained in the previous section the values of $G^2$ and
$F_h$ are unknown in MMRN. So we will used theoretical ratios to eliminate 
the dependence on these parameters  to compare our expressions with 
experimental results. 
We will now write down the theoretical expressions that  can be directly 
compared to the experimental data found in Table~\ref{tab1}.

Using Eqs.~(\ref{def3}) -- (\ref{def41}) we obtain  
\begin{equation}  
\left(\frac{m_{\mu}}{m_{\tau}}\right)^5 \frac{\Gamma^{\tau e}}  
{\Gamma^{\mu e}}= \frac{{\cal R}^{\tau}f^{\tau  
e}(x,y,\delta_{e\tau},\delta_{P\tau},\delta_{F\tau})}{{\cal R}^{\mu}
f^{\mu e}(x,y,\delta_{e\mu},\delta_{P\mu},\delta_{F\mu})}=
\left(\frac{m_{\mu}}{m_{\tau}}\right)^5
\frac{B^{\tau e}\tau_{\mu}}{B^{\mu e}\tau_{\tau}} \equiv
\left(\frac{G_\tau}{G_\mu}\right)^2 ,
\label{tu}
\end{equation}  
with $\tau_{\tau}$ and $\tau_{\mu}$  being respectively the tau and the 
muon lifetimes, $B^{l'l}$ the branching ratio for the decay  
$l' \to l \bar \nu_l \nu_{l'}$ and
  
\begin{equation}  
\frac{\Gamma^{\tau \mu}}  
{\Gamma^{\tau e}}=
\frac{f^{\tau  
\mu}(x,y,\delta_{\mu\tau},\delta_{P\tau},\delta_{F\tau})}{f^{\tau e}  
(x,y,\delta_{e\tau},\delta_{P\tau},\delta_{F\tau})}=
\frac{B^{\tau \mu}}{B^{\tau e}}.
\label{taufra}
\end{equation}

From Eqs.~(\ref{hadw}), (\ref{pimu}) and (\ref{pie}) we obtain for the pion 
decays 
\begin{equation}
\frac{\Gamma^{ \pi e}}{\Gamma^{\pi \mu}}=
\frac{{\cal R}_{\pi e} f^{\pi e}(x,y,\delta_{\pi e},\delta_{P e},\delta_{F e})}
{{\cal R}_{\pi\mu}
f^{\pi \mu}(x,y,\delta_{\pi \mu},\delta_{P \mu},\delta_{F \mu})}
=\frac{B^{\pi e}}{B^{\pi\mu}},
\label{piemu}  
\end{equation} 
where  $B^{\pi l}$ is the branching ratio for the decay  
$\pi \to l \nu_l $ ($l=\mu,e$).
For the kaon decays an alike expression can be derived. Before we give
this expression we would like to make some remarks.

Kaon leptonic decay measurements are not only less precise than the pion 
leptonic decay ones but also suffer from an important background 
contamination. 
The average leptonic width given in PDG is dominated by the result of one
experiment, the CERN-Heidelberg experiment\cite{he1,he2}. 
In order to avoid  the contamination of 
$K_{l2}$ ($K^{+}\rightarrow l^{+}\nu _l$) events 
by beta decay  $K_{l3}$ $(K^{+}\rightarrow l^{+}\nu _l\pi ^0)$ events, 
experimentalists are forced  to impose a cut in the measured
momentum of the final charged lepton.
For massless neutrinos in $K_{l2}$ decays 
one expects the momentum $p_l$ ($l=e,\mu$), to be monochromatic  
i.e., $p_e=247$~MeV for the electron channel and $p_\mu =236$~MeV 
for the muon channel. 
Based on this,  $K_{e2}$ events  are experimentally characterizes  as 
having  $240$~MeV $\leq p_e\leq 260$~MeV and 
$K_{\mu2}$ events  as having $220$~MeV $\leq p_\mu \leq 252$~MeV~\cite{he1,he2}.

If neutrinos produced in these decays are massive we expected
as many lines in the spectrum of charged lepton as the number of
massive neutrinos. For a massive neutrino with mass $m_i$

\[
m_K=\sqrt{p_l(m_i)^2+m_l^2}+\sqrt{p_l(m_i)^2+m_i^2},
\]
which can be solved in terms of the final lepton momentum, $p_l(m_i)$, 
giving~\cite{winter}
\begin{equation}
p_l(m_i)=p_l(0)\sqrt{1-\frac{2\left( m_K^2+m_l^2\right) m_i^2-m_i^4}
{4m_K^2p_l(0)^2}}, 
\label{p_i}
\end{equation}
where $m_l$ is the mass of the charged lepton and 
$m_K$ is the mass of the kaon and $p_l(0)$ is the momentum for a 
massless neutrino $p_l(0)=\displaystyle\frac{m_K^2-m_l^2}{2m_K}$.

The experimental lower cut in the momentum of the final lepton 
together with Eq.(\ref{p_i}) imply a maximum value for the observable 
neutrino mass~\cite{ro}. Explicitly for $p_e~>~240$~MeV we have 
$m_i<m_{e}^{cut}=82$~MeV and for 
$p_{\mu}>220$~MeV, $m_i<m_{\mu}^{cut}=118$~MeV. That means, neutrinos
with a mass greater than $118$~MeV are not visible in either of
these decays.

These restrictions imply that Eqs.~(\ref{pimu}) and (\ref{pie}) 
will have to be changed for the kaon case : 
$f^{K e} \rightarrow \hat f^{K e}$  where

\begin{equation}
\hat f^{K e}(x,y,\delta_{K e},\delta_{P e},\delta_{F e}) 
=(1-x) \Gamma^{K e}_1+ 
(1-z) x \Gamma^{K e}_P\theta(m_{e}^{cut}-m_P) 
+z x \Gamma^{K e}_F\theta(m_{e}^{cut}-m_F),
\label{ke}  
\end{equation}
and also 
$f^{K \mu} \rightarrow \hat f^{K \mu}$ where

\begin{eqnarray}
\hat f^{K \mu}(x,y,\delta_{K \mu},\delta_{P \mu},\delta_{F \mu})
&=&(yx+1-y)\Gamma^{K \mu}_1
+(1-z) (1-x) y \Gamma^{K\mu}_P \theta(m_{\mu}^{cut}-m_P) \nonumber \\
&+&y(1-x)z \Gamma^{K \mu}_F \theta(m_{\mu}^{cut}-m_F),
\label{kmu}  
\end{eqnarray}  
so that finally  we have
\begin{equation}
\frac{\Gamma^{ K e}}{\Gamma^{K \mu}}=
\frac{{\cal R}_{K e} \hat f^{K e}(x,y,\delta_{K e},\delta_{P e},\delta_{F e})}
{{\cal R}_{K \mu}
\hat f^{K \mu}(x,y,\delta_{K \mu},\delta_{P \mu},\delta_{F \mu})}
=\frac{B^{K e}}{B^{K \mu}},
\label{kemu}  
\end{equation}  
where  $B^{K l}$ is the branching ratio for the decay 
$K \to l \nu_l $ ($l=\mu,e$).

For the $Z^0$ invisible width we  use

\begin{equation}
\Gamma^{\mbox{\scriptsize inv}}(Z \rightarrow \nu's)
=\sqrt{\frac{\Gamma_{11}^{\mu e} G_\mu^2}{ f^{\mu
e}(x,y,\delta_{e\mu},\delta_{P \mu},\delta_{F \mu})}}
\Gamma_0^{\mbox{\scriptsize SM}} 
 (2+(1-z^2) \chi_{PP}+2(1-z)z \chi_{PF} 
+z^2 \chi_{FF}).  
\label{Zwidth}  
\end{equation}

Now to establish the allowed regions for the free parameters of MMRN we have 
built the $\chi^2$ function
\begin{equation}
\chi^2(x,y,m_P,m_F) = \sum_{i=1,5}\frac{(F_i-F_i^{\text{exp}})^2}{\sigma_i^2}
\label{Xi2},
\end{equation}
\noindent
where each $F_i$ is the theoretical value calculated using one of the   
expressions given in Eqs.\ (\ref{tu}),(\ref{taufra}),(\ref{piemu}),
(\ref{kemu}) and (\ref{Zwidth}), and $F_i^{\text{exp}}$ 
and $\sigma_i$ are its corresponding 
experimental value and error according to Table~\ref{tab1}.

We have minimized this $\chi^2$ function with respect to its four parameters. 
The minimum $\chi^2$ found for one d.o.f. (five experimental data points 
minus four  free parameters) is  $\chi^2_{\mbox{\scriptsize min}} = $ 1.29 
for $x=.22~10^{-5}$, $y= 0.47$, $m_P=.28$ MeV and $m_F=1.10$ MeV,
this is a bit smaller than  $\chi^2_{\mbox{\scriptsize SM}} = $ 1.33, that 
we get  for $x=y=z=0$. The error matrix corresponding to the result of our 
minimization is: 

\begin{equation}  
\left(  
\begin{array}{cccc}  
V_{m_P m_P} & V_{m_P m_F} & V_{m_P x} & V_{m_P y} \\
V_{m_F m_P} & V_{m_F m_F} & V_{m_F x} & V_{m_F y} \\
V_{x m_P} & V_{x m_F} & V_{x x} & V_{x y} \\
V_{y m_P} & V_{y m_F} & V_{y x} & V_{y y} \\
\end{array}\right)
=\left(  
\begin{array}{cccc}  
.69~10^{-7} & .51~10^{-5} & .15~10^{-9} & 0. \\  
.51~10^{-5} & .43~10^{-3} & .12~10^{-7} & 0. \\   
.15~10^{-9} & .12~10^{-7} & .72~10^{-12} & 0. \\
0.          & 0.          & 0.           & .36~10^{-11} \\
\end{array}\right).  
\label{errmat}
\end{equation}

\begin{figure}
\begin{center}
\mbox{\qquad\epsfig{file=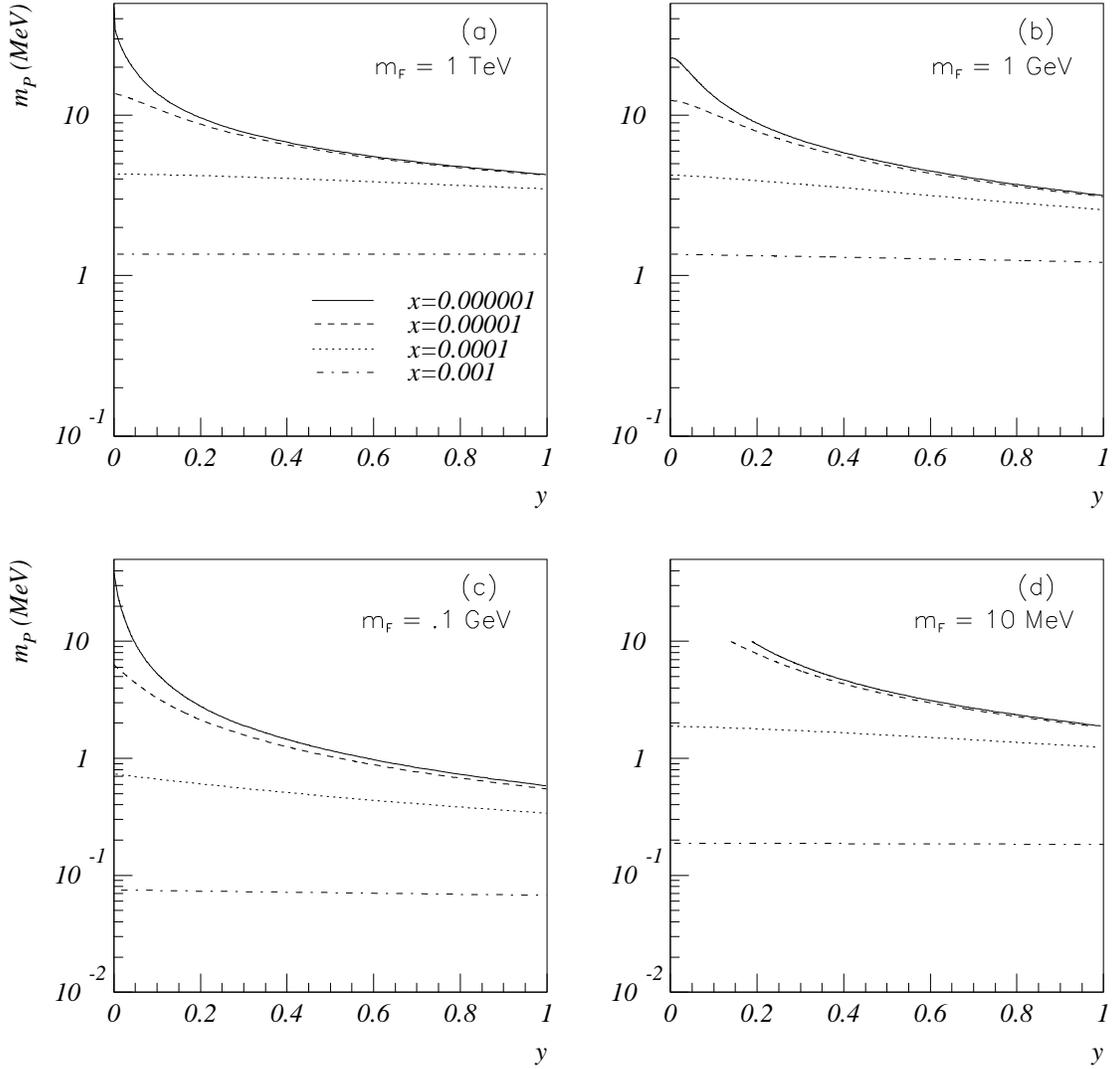,width=\linewidth}}
\end{center}
\caption{
Bellow each of the displayed curves, for a fixed value of $x$, 
we have the allowed region in the plane $m_F \times y$, at 90\% C.\ L.\, 
for  (a) $m_F=$ 1 TeV, (b) $m_F=$ 1 GeV, (c) $m_F=$ .1 GeV 
and (d) $m_F=$ 10 MeV.}
\label{fig1}
\end{figure}

We have computed the 90\% C.L.\ contours determined by 
the condition $  \chi^2 = \chi^2_{\mbox{\scriptsize min}} + 7.78 $. 
In order to display our results we have fixed the values of $m_F$ and 
presented the allowed regions in a $m_P \times y$ plot for several values 
of $x$. We have chosen to display the allowed regions for four different 
$m_F$ values to give an idea of the general behavior. 
This is shown in Figs.\ \ref{fig1}.

We note that our  $\chi^2$ function is very sensitive to changes in $x$ 
and $m_P$ but rather not so sensitive to $y$ or $m_F$. This behavior reflects 
on the fact that the maximum possible value of $m_P$ for each contour 
we have obtained, reached at $y \to 0$, is very sensitive to $x$ but not so 
sensitive to $m_F$. 
For $x>10^{-4}$  we see that the maximum allowed $m_P$ depends on $m_F$ 
but is almost independent of  $y$. In fact,  
this is expected as all our expressions become independent 
of $y$ as $x \rightarrow 1$. 
The absolute maximum allowed value of $m_P$,  
for $x,y \to 0 $, consistent with the data is $\simeq$  40 MeV. 
This is still true even if $m_F>$ 1 TeV.

We observe that the contours in the $m_P \times y$ plane have basically 
the same shape and allow for a lower maximum  value of $m_P$ as a function 
of $y$ and as $m_F$ decreases. Nevertheless there are two values for $m_F$ 
that change the behavior of the allowed contours. This is due to the fact 
that  the presence of massive neutrinos in the considered decays depends 
on kinematical constraints.
At $m_F = m_K-m_e$  higher values of $m_P$ as a function of $y$ become 
possible, here $m_F$ starts to participate in kaon decays. At 
this point the contour curve changes a little bit its shape  and becomes less 
restrictive. From then on, as $m_F$ decreases, the allowed curves 
share once more the same shape and start again to constrain the parameters.
At  $m_F = m_\pi-m_\mu$ we have a new change of behavior and higher values of  
$m_P$ become allowed since now  $m_F$ can participate of all pion 
decays.  Again after that for smaller values of  $m_F$ the curves 
will confine even more the parameters.

In Fig.\ \ref{fig1}(a) we see bellow each one of the curves the allowed 
regions, at 90\% C.L., 
of $m_P$ as a function of $y$ for $m_F=$ 1 TeV and four different 
values of $x$. 
In Fig.\ \ref{fig1}(b) we see the same contours  for $m_F=$ 1 GeV. We note 
that the allowed regions are not much more limited than in the previous 
case even though we have decreased $m_F$ by three orders of magnitude.
In Fig.\ \ref{fig1}(c) we see the allowed contours for $m_F=$ .1 GeV. Here 
we have already passed by  $m_F = m_K-m_e$ where the first change in 
behavior occurred. 
Finally in Fig.\ \ref{fig1}(d) we see the allowed contours for 
$m_F=$ 10 MeV. Some comments are in order here. One can see that the allowed 
regions in this case, although $m_F$ is much smaller than in 
Fig.\ \ref{fig1}(c)  are less restrictive. This is because we have crossed 
the value $m_F = m_\pi-m_\mu$ as explained above. 
Note also that for the lowest values of $x$ the curves are interrupted 
by the condition that  $m_P \leq m_F$, this means that for $y \lesssim 0.15$  
the only prerequisite is  $m_P \leq m_F$.

For $10^{-2} \leq x \leq 1$ the maximum allowed $m_P$ is really independent 
of $y$. This case can be subdivided into three regions: 
(i) for $m_F >$ 495 MeV, $m_P^{\mbox{\scriptsize max}}$ is also 
independent of  $m_F$ as can be seen in Table~\ref{tab2}; 
(ii) for smaller values of $m_F$ the product 
$m_P^{\mbox{\scriptsize max}} \times x$ is constant with $m_F$ as shown 
in Table~\ref{tab3} and (iii) for $m_F<$ 43 keV there is no restriction  
on $x$ and $y$ for $m_P \leq m_F$. 

Note that our analysis was done in the context of a specific model and 
that we did not impose the ad hoc limit to neutrino masses 
used in Ref.~\cite{Ng}.

Some general remarks about our results are in order here. 
The $Z^0$ invisible width measurement at LEP along with the pion decay 
data were by far the most significant experimental constraints to the model 
parameters. The invisible width is today a extremely precise measurement and 
as one should expect imposes great restrictions on neutrinos couplings. 
The pion decay measurements are also very precise and being phase space 
limited two body decays they have great power in constraining neutrino 
masses and couplings as long as they can participate in pion decays. 
On the other hand the kaon decay and the lepton decay data we have analyzed 
have not been so effective in constraining the model. Kaon decays  
unfortunately suffer from experimental contamination which makes their data 
less useful at the present moment than one should hope it to be.
We would expect that experimental improvements  here would affect 
our results. The $\mu$ and $\tau$ lepton decays are three body decays 
containing two neutrinos in the final state. This explain the fact that 
although the experimental measurements are quite accurate the overall 
effect of these data is not so constrictive to masses and couplings of 
individual neutrinos.

\section{Neutrinoless double-$\beta$ decay}
\label{sec:dbeta}

Besides the experimental limits already imposed by the decays in the previous 
section, since our neutrinos have Majorana nature, we can hope to further 
restrict the mixing parameters of the model by imposing the constraint 
coming from the non observation  of neutrinoless double-$\beta$ decays, 
i.e. (A,Z)$\to $(A,Z+2) $+~ 2 e^-$  transitions. 
This type of  process can be analyzed in terms of an effective neutrino 
mass  $ \langle m_\nu \rangle$ given in MMRN by~\cite{doi}

\begin{equation}
\langle m_\nu \rangle =\sum_{i=P,F} (\Phi R)_{ei}^2 m_i F(m_i,A),  
\label{mnu}
\end{equation}
where $F(m_i,A)$ is the matrix element for the nuclear transition which is a 
function of the neutrino mass $m_i$.
This has been computed in the literature for a number of different 
nuclei as the ratio~\cite{klapdor} 

\begin{equation}
F(m_i,A) = \frac{M_{GT}(m_i)-M_F(m_i)}{M_{GT}(0)-M_F(0)}.
\label{rmj}
\end{equation}

The best experimental limit on neutrinoless double-$\beta$ decay 
comes from the observation of the  nuclear transition 
$^{76}$Ge~$\rightarrow ^{76}$Se.
The result of the calculation of the nuclear matrix element $F(m_i,A)$ for 
$^{76}$Ge~$\rightarrow ^{76}$Se transitions can be found in 
Ref.~\cite{klapdor} and we will now refer to this simply as $F(m_i)$. 
This ratio is unity for $m_i \lesssim $ 40 MeV.
For 40 MeV $ < m_i < $ 1 GeV we have used the following  parabolic fit
that agrees with Fig.\ 8 of Ref.~\cite{klapdor} up to less than 10 \%

\begin{equation}
\log F(m_i) =  -37.96 + 10.1 \log m_i -0.6719 (\log m_i)^2,
\label{pfit1}
\end{equation}
and for $m_i > $ 1 GeV one can use 

\begin{equation}
F(m_i) = 3.2 \, (10^8 \mbox{eV}/m_i)^2,
\label{pfit2}
\end{equation}
with $m_i$  in eV in both of the above expressions. 

We have used Eqs.\ (\ref{pfit1}) and (\ref{pfit2}) along with the current 
best experimental limit $ \left| \langle m_\nu \rangle \right| < 0.6$ eV 
at 90\% C.\ L.~\cite{hmbb}  to draw our conclusions about the possible 
extra constraints that might be imposed to our previous results.

Due to the behavior of the nuclear matrix element $F(m_i)$ in
$^{76}$Ge~$\rightarrow ^{76}$Se transitions  and taken into account 
our previous results which always exclude $m_P>$ 40 MeV,
we conclude that we have in MMRN three different regions to inspect:

\begin{description}
\item (a) $m_P$,$m_F \leq $ 40 MeV; 
\item (b) $m_P<$ 40 MeV and 40 MeV $< m_F< $ 1 GeV;
\item (c) $m_P<$ 40 MeV and $m_F \geq $ 1 GeV. 
\end{description}

In case (a) $F(m_P)=F(m_F)=1$ and Eq.(\ref{mnu}) gives  

\begin{equation}
\langle m_\nu \rangle = (\Phi R)_{eP}^2m_P+ (\Phi R)_{eF}^2m_F=s_\beta ^2\left(
-c_\alpha ^2m_P+s_\alpha ^2m_F\right) =0;  
\label{mnu1}
\end{equation}
here, it is clear, the mixing parameters cannot be further constrained by  
the neutrinoless double-$\beta$ decay limit.
In cases (b) and (c) we have $F(m_P)=1$ and 

\begin{equation}
\langle m_\nu \rangle =s_\beta ^2s_\alpha ^2m_F\left( F(m_F)-1\right)
=xz m_F\left( F(m_F)-1\right),   
\label{mnu2}
\end{equation}
so in these cases extra limits on the mixing parameters can be expected.

\begin{figure}
\begin{center}
\mbox{\qquad\epsfig{file=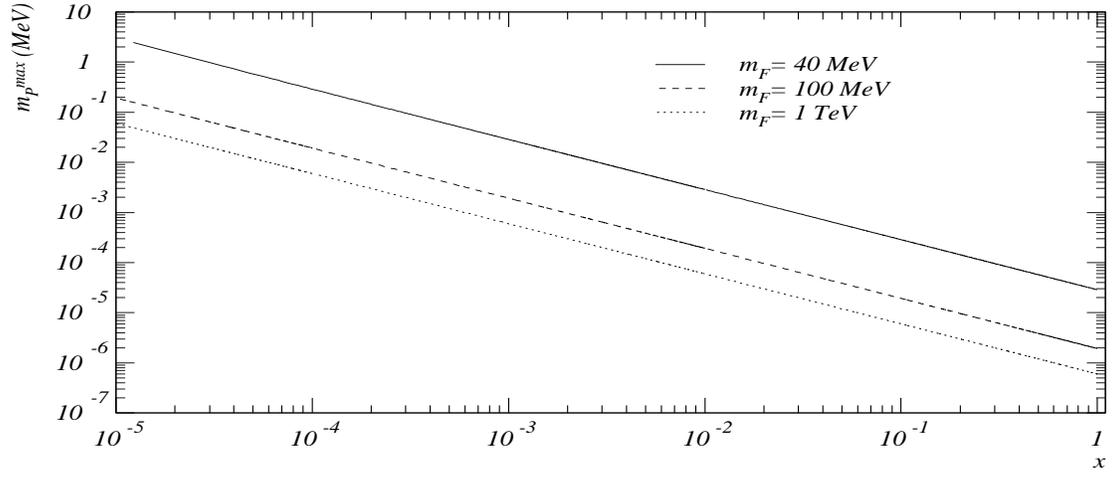,width=\linewidth,height=8cm}}
\end{center}
\caption{Maximum allowed value of $m_P$ as a function of $x$ 
for three different values of $m_F$ compatible with the 
neutrinoless double--$\beta$ decay limit.}
\label{fig2}
\end{figure}

Using Eq.\ (\ref{pfit1}) in Eq.\ (\ref{mnu2}) and imposing the current 
experimental limit of 0.6 eV one gets the maximum possible value 
of the product $xz$ allowed by the data. 
In region (c) we use Eq.\ (\ref{pfit2}) in  Eq.\ (\ref{mnu2}) and again impose 
the experimental limit. 
This procedure permits us to compute the maximum allowed value for $m_P$, 
$m_P^{\scriptsize \mbox{max}}$, as a function of $x$ for a given $m_F$. 
This can be seen in Fig.\ \ref{fig2} for three different values of 
$m_F$. 

For example in region (c), for $m_F=$ 1 TeV and $x \sim 10^{-5}$, 
$m_P \lesssim 0.06$ MeV. In region (b)  for $m_F=$ .1 GeV and 
$x \sim 10^{-5}$, $m_P \lesssim 0.2$ MeV. Both results  are independent of 
the values of $y$. 
For higher values of $x$ the limits on $m_P$ are even more strict.
We see from this that in regions (b) and (c) the neutrinoless
double-$\beta$ decay limit can severely constrain the parameters of 
the model.

\section{Conclusions}  
\label{sec:conc}  

We have analyzed the constraints imposed by recent experimental data 
from $\mu$ decay, $\tau$ , $\pi$ and $K$ leptonic decays, the 
$Z^0$ invisible width on the values of the four mixing parameters,
$x$, $y$, $m_P$ and $m_F$, of the MMRN model. 

We have found regions allowed by the combined data at 90\% C.\ L.\ 
in the four parameter space. These allowed regions are very sensitive to 
changes in the values of $x$ and not so sensitive to changes in $y$.
We were also able to find that the maximum possible  value 
for the lightest neutrino mass $m_P$, obtained in the limit $x,y \to 0$,
is about 40 MeV, even if $m_F>$ 1 TeV.
Although this is not so restrictive as the maximum value of $\nu_\tau$ 
obtained experimentally by ALEPH~\cite{pdg98} it is very interesting to see 
that the electroweak data alone can indirectly lead to a value already so 
limited.

We also have investigated and found that for $m_F>$ 40 MeV 
the most recent neutrinoless double-$\beta$ decay limit can constrain 
considerably  more the model free parameters, in particularly the maximum  
allowed value of $m_P$. For instance if $m_F= 1$ TeV and $x=1$, then  
$m_P^{\text{max}} \sim$ 0.6 eV.

After combining the results from the particle decay analysis with the 
constraints from  neutrinoless double-$\beta$ decay we get finally :

\begin{description}
\item (a) for $m_P$,$m_F \leq $ 40 MeV, the constraints on the free 
parameters are simply given by accelerator decay data, such as in 
Fig.\ \ref{fig1}(d); 
\item (b) for $m_F> $ 40 MeV, the limit from neutrinoless double-$\beta$ 
decay constrains the maximum value of $m_P$ to much smaller values than what 
are still possible with the accelerator data, as shown in Fig.\ \ref{fig2}.
\end{description}

We have not used the available data on charm (or even beauty) meson 
leptonic decay modes such as $D_s \rightarrow  \mu \nu_\mu$ and 
$D_s \rightarrow  \tau \nu_\tau$. This data have very large uncertainties 
attached to them and would not affect our results at the present moment.
We also have not used the data from $\tau \rightarrow \pi (3\pi) \nu_\tau$ 
due to the fact that they are experimentally less precise and 
theoretically more problematic than $\tau$ leptonic decays. We do not 
think these two modes would affect very much, if at all, our conclusions.


\section{Acknowledgments}

This work was supported by DGICYT grant PB95-1077, by the EEC 
under the TMR contract ERBFMRX-CT96-0090, by Conselho Nacional de 
Desenvolvimento Cient\'{\i}fico e Tecnol\'ogico (CNPq) and 
by Funda\c{c}\~ao de Amparo \`a Pesquisa do Estado de S\~ao Paulo (FAPESP).


\section*{APPENDIX : Radiative Correction formulae}

The leading radiative corrections to the lepton decay process 
$ {l'} \rightarrow l \bar\nu_l \nu_{l'}$, ${\cal R}^{l'}$, 
is give by~\cite{sirlin86}

\begin{equation}
{\cal R}^{l'} = \left[1+\frac{\alpha(m_{l'})}{2\pi}\left(\frac{25}{4}-\pi^2\right)\right]\left(1+\frac{3m^2_{l'}}{5m_W^2}\right),
\label{leprad}
\end{equation} 
where $m_{l'}$ is the initial lepton mass, $ m_W$ is the $W$ boson mass 
and $\alpha(m_{l'})$ is the running electromagnetic coupling constant.

The leading radiative corrections to hadron leptonic decays ${\cal R}_{hl}$ 
is given by~\cite{pdg98,sirlin} 

\begin{eqnarray}
{\cal R}_{h l} &=&  
\left[ 1 + \frac{2 \alpha}{\pi}  
\ln\left(\frac{M_Z}{m_\rho}\right)\right]   
\left[1+\frac{\alpha}{\pi} F(\delta _{l h}) \right]    
\nonumber       \\  
& & \times \left\{ 1 - \frac{\alpha}{\pi}  
\left[\frac{3}{2}\ln\left(\frac{m_\rho}{m_h}\right) +  
C_1 +   
C_2 \frac{m_l^2}{m_\rho^2} \ln\left(\frac{m_\rho^2}{m_l^2}\right) +  
C_3 \frac{m_l^2}{m_\rho^2} + \dots  
\right]\right\},
\label{pirc}  
\end{eqnarray}  
where 

\begin{eqnarray}  
F(x) &=&  
3\ln x + \frac{13-19 x^2}{8(1-x^2)} -   
\frac{8-5x^2}{2(1-x^2)^2}x^2\ln x   
\nonumber \\  
& &  
\mbox{} - 2 \left( \frac{1+x^2}{1-x^2}\ln x +1 \right)\ln(1-x^2) +  
2\left( \frac{1+x^2}{1-x^2} \right)L(1-x^2).
\end{eqnarray}  
Here, $m_\rho = 796$~MeV is  
the $\rho$ meson mass, $M_Z$  the $Z^0$ boson mass, $\alpha$ is the fine   
structure constant and $m_{l}$ is the final lepton mass.
$C_i$ are structure constants whose numerical value have 
large uncertainties and for this reason these terms will be neglected by us~\cite{pdg98}.Also, in the above,  
$L(z)$ is defined by  
  
\begin{equation}  
L(z) = \int_0^z \frac{\ln(1-t)}{t} dt.  
\end{equation}  



\begin{table}[h]
\begin{center}
\begin{tabular}{|l|c|}
\hline
 & Based on PDG 1998 Data \\
\hline
$m_\tau$  & $1777.05^{+0.29}_{-0.26}$ MeV \\
\hline
$m_\mu$  & $105.658389\pm0.000034$ MeV \\
\hline
$m_e$ &  $0.51099907\pm0.00000015$ MeV \\ 
\hline
$m_\pi$ & $139.56995\pm0.00035$ MeV\\ 
\hline
$m_K$ & $493.677\pm0.016$ MeV \\
\hline
$m_W$ & $80.41\pm0.10$ GeV \\
\hline
$\tau_\tau$ & $290.0\pm1.2~10^{-15}$ s\\
\hline
$\tau_\mu$ & $(2.19703\pm0.00004)~10^{-6}$ s\\
\hline
$\tau_\pi$ & $(2.6033\pm0.0005)~10^{-8}$ s\\
\hline
$\tau_K$ & $(1.2386\pm0.0024)~10^{-8}$ s\\
\hline
$B^{\tau \mu}$ & $17.37\pm0.09$ \\
\hline
${B^{\tau e}}$ & $17.81\pm0.07$ \\
\hline
${B^{\pi e}}$ & $(1.230\pm0.004)10^{-4}$ \\
\hline
${B^{\pi \mu}}$ & $(99.98770\pm0.00004)10^{-2}$ \\
\hline
${B^{K e}}$ & $(1.55\pm0.07)10^{-5}$ \\
\hline
${B^{K \mu}}$ & $(63.51\pm0.18)10^{-2}$ \\
\hline
$(\displaystyle\frac{G_\tau}{G_\mu})^2$ & $1.0027\pm0.0089$ \\
\hline
$\displaystyle\frac{B^{\tau \mu}}{B^{\tau e}}$ & $0.9753\pm0.0089$ \\
\hline
$\displaystyle\frac{B^{\pi e}}{B^{\pi \mu}}$ & $(1.2302\pm0.004) 10^{-4}$ \\
\hline
$\displaystyle\frac{B^{K e}}{B^{K \mu}}$ & $(2.4406\pm0.1171) 10^{-5}$ \\
\hline
$\Gamma^{\mbox{\scriptsize inv}}(Z\rightarrow \nu's)$ & $500.1 \pm 1.8$ MeV $^{(\star)}$ \\
\hline
\end{tabular}
\end{center}
\protect \caption{Experimental values and ratios used to constrain the mixing 
parameters. $^{(\star)}$ This value of $\Gamma^{\mbox{\scriptsize inv}}$ was actually taken from Ref.\ [20].}
\label{tab1}
\end{table}


\begin{table}[h]
\begin{center}
\begin{tabular}{|c|c|}
\hline
$x$ & $m_P^{\mbox{\scriptsize max}}~$ (MeV) \\
\hline
\hline
$1$    & 4.3 $10^{-2}$ \\
$10^{-1}$  & 1.3 $10^{-1}$ \\
$10^{-2}$ & 4.3 $10^{-1}$ \\
\hline
\end{tabular}
\end{center}
\protect \caption{ Values of $m_P^{\mbox{\scriptsize max}}$ for $m_F\geq$ 495 MeV
 and $10^{-2} \leq x \leq 1$. }
\label{tab2}
\end{table}

\begin{table}[h]
\begin{center}
\begin{tabular}{|c|c|}
\hline
$m_F~$ (MeV) & $m_P^{\mbox{\scriptsize max}} \times x~$ (MeV) \\
\hline
\hline
$100$  & 7.5  $10^{-5}$  \\
$35$   & 6.05 $10^{-5}$ \\
$10$   & 1.88 $10^{-4}$ \\
$1$    & 1.87 $10^{-3}$ \\
$0.1$  & 1.87 $10^{-2}$ \\
\hline
\end{tabular}
\end{center}
\protect \caption{ Values of $m_P^{\mbox{\scriptsize max}} \times x$ for 
$m_F\leq$ 100 MeV and $10^{-2} \leq x \leq 1$. }
\label{tab3}
\end{table} 



\end{document}